\documentclass[aps,prl,twocolumn,showpacs,superscriptaddress]{revtex4-1}  
\usepackage{graphicx}  
\usepackage{dcolumn}   
\usepackage{bm}        
\usepackage{amssymb}   

\begin{document}

\title{Competing Ferri- and Antiferromagnetic Phases in Geometrically Frustrated LuFe$_{2}$O$_{4}$}

\author{J. de Groot}
\affiliation{Peter Gr\"{u}nberg Institut PGI and J\"{u}lich Centre for Neutron Science JCNS, JARA-FIT, Forschungszentrum J\"{u}lich GmbH, 52425 J\"{u}lich, Germany}

\author{K. Marty}
\affiliation{Oak Ridge National Laboratory, Oak Ridge, Tennessee 37831, USA}

\author{M.$\,$D. Lumsden}
\affiliation{Oak Ridge National Laboratory, Oak Ridge, Tennessee 37831, USA}

\author{A.$\,$D. Christianson}
\affiliation{Oak Ridge National Laboratory, Oak Ridge, Tennessee 37831, USA}

\author{S.$\,$E. Nagler}
\affiliation{Oak Ridge National Laboratory, Oak Ridge, Tennessee 37831, USA}

\author{S. Adiga}
\affiliation{Peter Gr\"{u}nberg Institut PGI and J\"{u}lich Centre for Neutron Science JCNS, JARA-FIT, Forschungszentrum J\"{u}lich GmbH, 52425 J\"{u}lich, Germany}

\author{W.$\,$J.$\,$H. Borghols}
\affiliation{J\"{u}lich Centre for Neutron Science JCNS, Forschungszentrum J\"{u}lich GmbH, 52425 J\"{u}lich, Germany}
\affiliation{JCNS Outstation at FRM II, D-85747 Garching, Germany}

\author{K. Schmalzl}
\affiliation{J\"{u}lich Centre for Neutron Science JCNS, Forschungszentrum J\"{u}lich GmbH, 52425 J\"{u}lich, Germany}
\affiliation{JCNS Outstation at ILL, BP 156, 38042 Grenoble, France}

\author{Z. Yamani}
\affiliation{National Research Council, Canadian Neutron Beam Center, Chalk River, Ontario, Canada}

\author{S.$\,$R. Bland}
\affiliation{Department of Physics, Durham University, South Road, Durham, DH1 3LE, United Kingdom}

\author{R. de Souza}
\affiliation{Swiss Light Source, Paul Scherrer Institut, 5232 Villigen PSI, Switzerland}

\author{U. Staub}
\affiliation{Swiss Light Source, Paul Scherrer Institut, 5232 Villigen PSI, Switzerland}

\author{W. Schweika}
\affiliation{Peter Gr\"{u}nberg Institut PGI and J\"{u}lich Centre for Neutron Science JCNS, JARA-FIT, Forschungszentrum J\"{u}lich GmbH, 52425 J\"{u}lich, Germany}

\author{Y. Su}
\affiliation{J\"{u}lich Centre for Neutron Science JCNS, Forschungszentrum J\"{u}lich GmbH, 52425 J\"{u}lich, Germany}
\affiliation{JCNS Outstation at FRM II, D-85747 Garching, Germany}

\author{M. Angst}
\email{M.Angst@fz-juelich.de}
\affiliation{Peter Gr\"{u}nberg Institut PGI and J\"{u}lich Centre for Neutron Science JCNS, JARA-FIT, Forschungszentrum J\"{u}lich GmbH, 52425 J\"{u}lich, Germany}

\date{\today}

\begin{abstract}
We present a detailed study of magnetism in LuFe$_{2}$O$_{4}$, combining magnetization measurements with neutron and soft x-ray diffraction. The magnetic phase diagram in the vicinity of $T_{N}$ involves a metamagnetic transition separating an antiferro- and a ferrimagnetic phase. For both phases the spin structure is refined by neutron diffraction. Observed diffuse magnetic scattering far above $T_{N}$ is explained in terms of near degeneracy of the magnetic phases.
\end{abstract}
\maketitle
Magnetoelectric multiferroics are of interest for novel storage devices \cite{BIBES,ERENSTEIN}. LuFe$_{2}$O$_{4}$ was proposed to be a multiferroic with a novel mechanism for ferroelectricity, based on Fe$^{2+}\!/$Fe$^{3+}$charge order (CO) \cite{Ikeda} below $T_{CO}$$\sim$320$\,$K. Mainly for this reason, but also due to unrelated effects such as giant coercivity, it is currently attracting a lot of attention \cite{Angst1,Mulders2,XU,REN,XU2,ROUQUETTE,MULDERS,XIANG,SUBMANIAN,HARRIS,d0det,WU,WEN09,WEN10,WANG,XMCD}. Both charge and spin degrees of freedom are localized at the Fe sites, which are contained in triangular Fe-O bilayers, a highly frustrated arrangement. For the CO, competing instabilities suggested by diffuse scattering above $T_{CO}$ \cite{Angst1}, were indeed linked to geometrical frustration \cite{HARRIS}. Similar geometrical frustration effects can also be expected for the magnetism, the elucidation of which is important for understanding the magnetoelectric coupling and other intriguing effects such as giant coercivity \cite{WU}.\par

LuFe$_{2}$O$_{4}$ typically exhibits magnetic order or freezing below about 220 - 240$\,$K. There is consensus that the Fe spins have a strong preference to be aligned $||\,c_{\,\mathrm{hex}}$, perpendicular to the layers \cite{XIANG,d0det,XMCD,WU,IIDA86,WANG,IIDA}. The magnetic behavior thus arises from Ising-spins on triangular lattices. Consistent with the highly frustrated arrangement many unusual effects have been observed in different samples, including various cluster or spin glass states \cite{WANG,WEN09}, a magneto-structural transition at $T_{LT}$$\sim$170$\,$K \cite{XU,d0det} and an anomalous ``field-heating-effect'' \cite{IIDA86}. Strong sample-to-sample variations in magnetic behavior are found, attributed to tiny variations in oxygen stoichiometry. Despite the high current interest, the details of the magnetic field ($H$) - temperature ($T$) phase diagram underlying these unusual behaviors have not yet been established.\par

\begin{figure}[b]
\includegraphics[width=0.99 \linewidth]{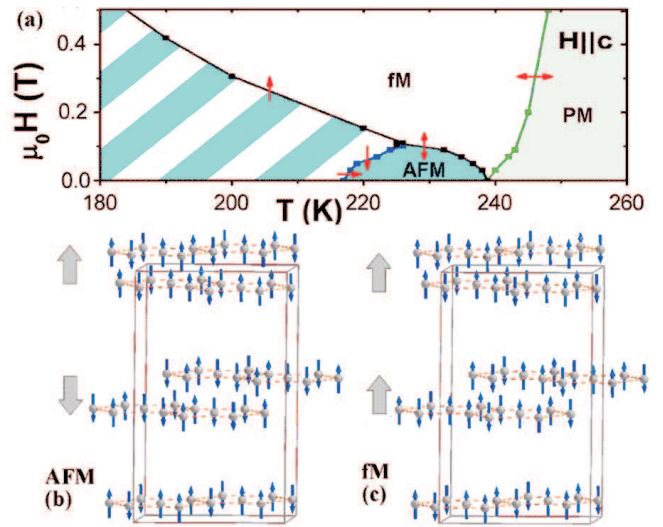}
\caption{\label{fig:Phaselow} (Color online) (a) Magnetic field $H$ - temperature $T$ phase diagram, which exhibits a paramagnetic (PM) an antiferromagnetic (AFM) and a ferrimagnetic (fM) phase, extracted from various $M(H)$ and $M(T)$ curves. The hysteretic region where either fM or AFM can be stabilized is hatched. Arrows across phase lines indicate for which measurement direction it is observed given the hysteresis. Spin structure in $C2/m$ cell \cite{SUPPLEMENT} of the AFM (b) and fM phase (c). Grey arrows indicate bilayer net magnetization.}
\end{figure}

\begin{figure}
\includegraphics[width=0.99\linewidth]{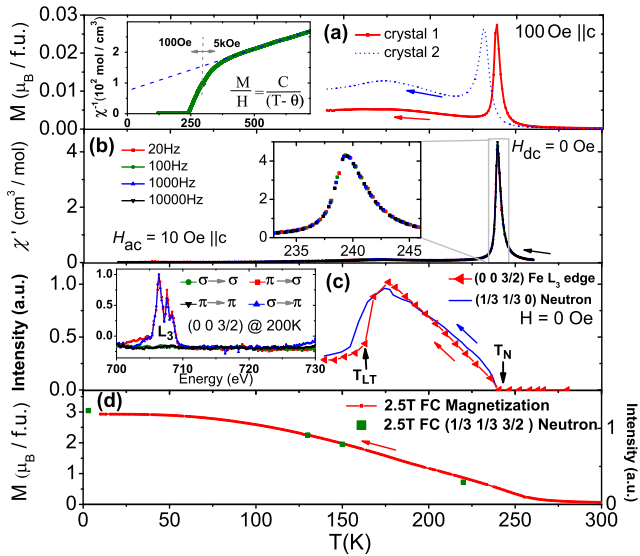}
\caption{\label{fig:ACMS} (Color online) $T$-dependence of various properties, all measured on cooling. (a) Magnetization $M$, compared with another sample. Inset: Inverse susceptibility $H$/$M$ with Curie-Weiss fit (dashed blue line) from 450$\,$K to 750$\,$K. (b) Ac-susceptibility measurement with different driving frequencies; the inset shows an enlarged area at $T_{N}$. (c) Integrated intensity of $(00\frac{3}{2})$ x-ray reflection at 706.4$\,$eV (Fe-$L_{3}$) and the $(\frac{1}{3}\frac{1}{3}0)$ neutron reflection, both in $H\!=\!0$. Inset: Energy scans across the Fe-L edges at $(00\frac{1}{2})$ with different incoming and outgoing polarization directions. (d) Integrated intensity for the $(\frac{1}{3}\frac{1}{3}\frac{3}{2})$ neutron reflection and $M(T)$, both in 2.5$\,$T.}
\end{figure}

In this work, we present a detailed study of the $H\!-\!T$ phase diagram above $T_{LT}$ of LuFe$_{2}$O$_{4}$ \cite{LOWPHASE} by magnetization and neutron and soft x-ray diffraction, revealing competing antiferromagnetic (AFM) and ferrimagnetic (fM) spin structures. The main focus is on samples with sharp magnetic transitions at $T_{N}\!\sim$240$\,$K to long-range spin order \cite{d0det}, which we propose to best approximate the intrinsic defect-free magnetic behavior of LuFe$_{2}$O$_{4}$. We demonstrate that at $T_{N}$ and $H\!=\!0$ fM and AFM instabilities, which correspond respectively to ferro and antiferro stacking of equivalently ordered bilayers, are nearly degenerate. These bear a striking resemblance with the two nearly degenerate CO instabilities \cite{Angst1,HARRIS} at $T_{CO}$, which we attribute to the similarity of binary (Ising spins or valence states) order emerging from competing interactions on the same strongly frustrated lattice. Diffuse magnetic scattering above $T_{N}$ indicates a random stacking of still individually ferrimagnetically ordered bilayers. We emphasize that although AFM-fM meta-magnetism has not been reported previously and may not be resolvable in the majority of LuFe$_{2}$O$_{4}$ samples, our results have strong implications for the general nature of magnetism in this material. In particular, our results underline the importance of geometrical frustration in LuFe$_{2}$O$_{4}$, both for charge and spin order.\par

We studied various LuFe$_{2}$O$_{4}$ single crystals from the same batch as in \cite{d0det,XU,XU2,Angst1}. Dc magnetization $M$ and ac susceptibility $\chi'$ measurements in $H||c_{\mathrm{hex}}$ were performed with commercial (Quantum Design) equipment. Polarized neutron diffraction in $H\!\sim\!0$ (except a small guide field less than 10$\,$Oe) was performed on DNS at FRM-II and non-polarized neutron diffraction with $H||c_{\mathrm{hex}}$ up to 2.5$\,$T on D23 at ILL and C5 at Chalk River Laboratories, all using the crystal labeled S2 in \cite{d0det}. Resonant x-ray diffraction at the Fe $L_{3}$ edge was performed on the SIM beamline (RESOXS endstation) at the SLS. For comparison with previous work all reflections have been indexed in hexagonal notation.\par

\begin{figure}
\includegraphics[width=0.99 \linewidth]{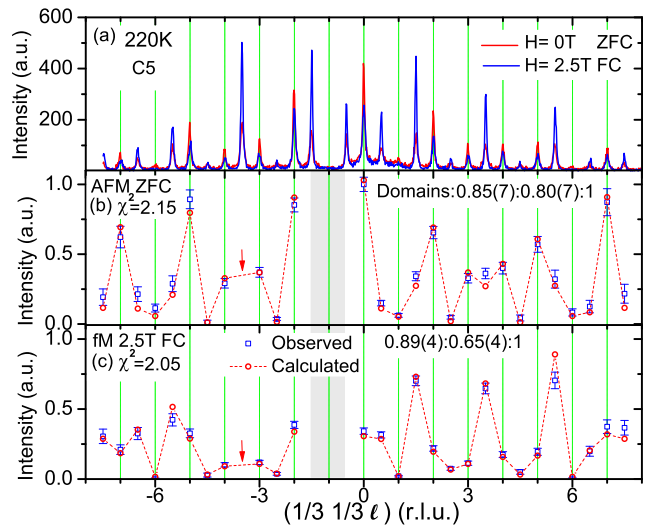}
\caption{\label{fig:TEST} (color online). (a) Neutron diffraction (N5) pattern at 220$\,$K along the $(\frac{1}{3}\frac{1}{3}\ell)$ line in $H\!=$2.5$\,$T and in $H\!=\!0$. (b,c) Integrated intensity for scans along $(\frac{1}{3}\frac{1}{3}\ell)$ in both magnetic phases corrected at 220$\,$K as described in the text. The solid red line represents the result from the model of \cite{d0det} for the spin structures shown in Fig.1. The gray area indicate the magnet dark angle and the red arrow a reflection affected by a second grain.}
\end{figure}

$M(T)$ measurements show variations of magnetic properties even among samples from one batch. One extreme exhibits characteristics matching those of \cite{PHAN}, with a strongly frequency-dependent peak in $\chi'$ around 225$\,$K indicating a transition into a glassy state. The other extreme exhibits at $T_{N}$$\sim$240K a sharp peak in $M$ and $\chi'$ without frequency-splitting  (Fig.\ref{fig:ACMS}a/b) and shows (Fig.\ref{fig:TEST}\,a) sharp magnetic reflections in neutron diffraction, indicating 3D long-range spin order rather than a glassy state. We also characterized samples with intermediate properties: the peak signifying magnetic ordering is shifted to lower $T$ (Fig.~\ref{fig:ACMS}a dotted line) and becomes weakly frequency-dependent. This indicates weakened magnetic correlations concomitant with ``glassiness'' and parasitic fM. In the following we focus on the type of samples showing the sharpest features in magnetization, where diffraction reveal sharp CO and magnetic reflections. Above $\sim$400$\,$K the inverse susceptibility $H/M$ (inset Fig.~\ref{fig:ACMS}a) follows a Curie-Weiss law with the effective moment $\mu_{\mathrm{eff}}$=5.51(9)$\mu_{B}$ expected for Fe$^{2+/3+}$ and a negative Weiss temperature of $\theta$=$-$307(9)$\,$K suggesting dominantly antiferromagnetic interactions, similar to YFe$_{2}$O$_{4}$ \cite{MATSUMOTO}.\par

\begin{figure}
\includegraphics[width=0.99 \linewidth]{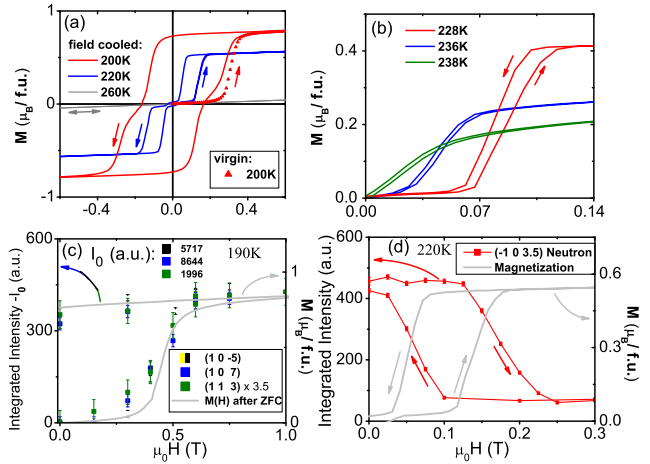}
\caption{\label{fig:HYST}(Color online) (a,b) $M(H)$ curves. The measurement direction is indicated by arrows. Virgin curves are measured after cooling in $H\!=\!0$. (c,d) Integrated intensity of different reflections in neutron diffraction (D23) vs. $H$; compared with $M(H)$, in (c) only the intensity change to $I_{0}$ on structural reflections is illustrated. }
\end{figure}

Isothermal magnetization $M(H)$ below $T_{N}$ (Fig.~\ref{fig:HYST}a/b) indicates a first-order metamagnetic transition, which becomes strongly hysteretic for lower $T$. The hysteretic region is indicated by the hatched area in Fig.~\ref{fig:Phaselow}a. The low-$T$ saturation moment of the high-$H$ phase (Fig.~\ref{fig:ACMS}d), is similar to previous findings \cite{IIDA,WU}, implying fM spin ordering. In contrast to this, the low-$H$ phase near $T_{N}$ seems to be AFM \cite{AFMPHASE}, with $M\!\propto\!H$ and no remanent moment.\par

An AFM phase at 220$\,$K and $H\!=\!0$ is inconsistent with the fM spin structure previously proposed \cite{d0det}. The absence of significant remanent $M$ at 220$\,$K could in principle be explained by the formation of compensating fM domains. However, the drastic effect of $H$ observed on several reflections with neutron diffraction, including a decrease of $(\frac{1}{3},\frac{1}{3},$integer$)$ and increase of $(\frac{1}{3},\frac{1}{3},$halfinteger$)$ reflections (Fig.~\ref{fig:HYST}a) and the emergence of additional intensity on structural reflections (Fig.~\ref{fig:HYST}c/d) show that the step in $M(H)$ clearly corresponds to a coherent effect, i.e., a genuine metamagnetic transition between two spin structures. The transition temperatures and fields from neutron scattering confirm the phase diagram from $M(H,T)$, including the large hysteresis.\par

The zero-field spin structure proposed in \cite{d0det} describes very well the $(\frac{1}{3},\frac{1}{3},$integer$)$ reflections, but do not account for newly observed magnetic reflections:
With soft x-ray diffraction we observed a sharp reflection at $(00\frac{3}{2})$ when the energy is tuned to the Fe $L_{3}$ edge. According to previous work the polarization analysis (inset Fig.~\ref{fig:ACMS}c) suggests that this is purely magnetic, resulting, as expected, from spins $||c_{\,\mathrm{hex}}$ \cite{STAUB}. The similar $T$-dependence to the $(\frac{1}{3}\frac{1}{3}0)$ reflection (Fig.~\ref{fig:ACMS}c) indicates that it originates from the same spin structure, further supported by the suppression of the equivalent $(\overline{1}02)\!+\!(00\frac{3}{2})$ reflection at the metamagnetic transition (Fig.~\ref{fig:HYST}d).\par

For the spin model in \cite{d0det} calculations show zero magnetic intensity for these reflections. This model therefore has to be excluded. This spin structure \cite{d0det} resulted from representation analysis based on the then only known $R\overline{3}m$ crystallographic cell with no CO and a single Fe site, leading to a very small number of spin structures to be considered. To describe the new observed magnetic reflections, we work within a 6$\times$ larger $C2/m$ CO cell \cite{XU2}, which corresponds to the magnetic cell for one domain according to all observed magnetic reflections.\par

We take the most expansive approach by ignoring symmetry and considering all $3^{12}$ possible spin configurations of 12 Fe Ising spins (allowing for partial disorder) in the primitive cell \cite{SUPPLEMENT}. Of these, $\sim$15000 structures yield the same relative intensities at $(\frac{1}{3},\frac{1}{3},$integer$)$ as the structure proposed in \cite{d0det}. To distinguish these structures, some broad-size restrictions for solutions can be made, based on the relative magnetic contribution of $\textbf{S}\!+\!(00\frac{3}{2})$ reflections and an upper limit of $(\frac{1}{3},\frac{1}{3},$halfinteger$)$ and structural $\textbf{S}\!+\!(000)$ reflections \cite{SUPPLEMENT}. These restrictions show that 7 symmetry-inequivalent spin structures can possibly be consistent with the observed magnetic diffraction in zero field. Refining these by fitting domain populations and a Debye-Waller factor as in \cite{d0det}, but including $(\frac{1}{3},\frac{1}{3},$halfinteger$)$, 6 models are rejected due to very large reduced $\chi^{2}$, the remaining solution is the AFM spin structure shown in Fig.~\ref{fig:Phaselow}b. In contrast to the rejected structures, this solution is fully ordered and has a simple relationship to the high-$H$ spin structure (see below). In this structure the spins of each bilayer are fM aligned ($\uparrow\uparrow\downarrow$), but the net moments of the bilayers are stacked antiferromagnetically, leading to the observed AFM behavior.\par

The moderate $\chi^{2}$ of 2.15 is due to systematically slightly higher intensities on $(\frac{1}{3},\frac{1}{3},$halfinteger$)$ reflections (Fig.~\ref{fig:TEST}b). A cause for this could be magnetic contrast due to different Fe$^{2+}$ and Fe$^{3+}$ moments. For the CO proposed in \cite{Angst1,Ikeda} no significant improvement on refinement with Fe$^{2+/3+}$ magnetic contrast is observed for all possible spin structures, but a CO configuration previously rejected due to charged bilayers \cite{Angst1} can further reduce $\chi^{2}$ to $\sim$ 1 for the above optimal spin structure. Furthermore, a $\chi^{2} \sim$1 can also be reached by considering a high-$H$ phase contamination, albeit with a 15\% phase fraction, which appears inconsistent with the remanent magnetization observed in Fig.~\ref{fig:HYST}d. Given the similar effects of cross-contamination and CO, the Fe$^{2+/3+}$ distribution cannot be conclusively \cite{SUPPLEMENT}.\par

The same approach was used for the high-$H$ phase \cite{SUPPLEMENT}. Comparing all $(\frac{1}{3}\frac{1}{3}\ell)$ with $\ell$ integer and halfinteger values according to the diffraction pattern in 2.5 $T$ (Fig.~\ref{fig:TEST}a) 245 possibilities remain. The $(\overline{1}02)\!+\!(00\frac{3}{2})$ reflection is strongly suppressed in the high-$H$ phase (Fig.~\ref{fig:HYST}d), reducing the possibilities to 42. After comparing the magnetic contribution on different structural reflections (Fig.~\ref{fig:HYST}c), only 18 solutions remain, corresponding to 3 symmetry inequivalent structures. Upon refinement, two are immediately rejected, the other (Fig.~\ref{fig:TEST}c) is presented in Fig.~\ref{fig:Phaselow}c. 

The fM solution is identical to the AFM solution, except that all Ising-spins in one Fe-O bilayer flip their sign, leading to the overall $\!2\!:\!1\!$ configuration of $\uparrow$ and $\downarrow$ spins consistent with the observed (Fig.~\ref{fig:ACMS}d) net moment. This different stacking of bilayer net magnetization between the AFM ($\uparrow\downarrow\uparrow\ldots$) and the fM ($\uparrow\uparrow\uparrow\ldots$) phase shares similarity with the competing CO instabilities at higher $T$. Phase competition and metamagnetic transitions between $\uparrow\downarrow\uparrow$ and $\uparrow\uparrow\uparrow$-stacking of net moments are expected for layered magnets with very strong Ising-anisotropy \cite{Stryjewski} and has been observed in a few model systems at low $T$, e.g. FeCl$_{2}$ \cite{Wilkinson}.

Intriguingly for LuFe$_2$O$_4$, in contrast to expectations in simple model systems, the AFM-fM transition extrapolates to $H\approx0$ for $T\rightarrow T_{N}$ as seen in Fig.~\ref{fig:HYST}b; i.e. at $T_{N}$ and $H\!=\!0$ the two phases seem to be essentially degenerate. The near-degeneracy of both charge and magnetic order is a hallmark of the importance of geometrical frustration in this system. The AFM/fM near-degeneracy in low $H$ can lead to parts of the sample being trapped in fM after cooling through $T_{N}$, particularly for samples with reduced $T_{N}$ (Fig.~\ref{fig:ACMS}a).\par

\begin{figure}
\includegraphics[width=0.99 \linewidth]{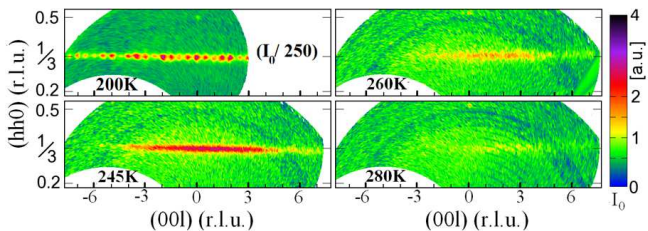}
\caption{\label{fig:DNS} (Color online) Reciprocal space map of the scattered intensity (logarithmic scale) in the$(hh\ell)$ plane by polarized neutrons in the spin flip channel (DNS) at different $T$.}
\end{figure}

The particular differences between the two nearly-degenerate spin-structures suggest that the intra-bilayer correlations are more dominant than the inter-bilayer correlations. Just above $T_{N}$ we may therefore expect a random stacking of the net moment of still medium-range ordered bilayers, i.e.\ a 2D-order \cite{FUNAHASHI}. In contrast to the ferrimagnetic ordered bilayers of LuFe$_2$O$_4$, for FeCl$_{2}$ the spins on triangular single layers are ferromagnetically coupled. For LuFe$_2$O$_4$, magnetic diffraction would result in strong diffuse scattering lines through $(\frac{1}{3}\frac{1}{3}\ell)$ above T$_{N}$, still reasonably sharp in-plane, but featureless along $\ell$. This is indeed observed, visible at 280$\,$K in Fig.~\ref{fig:DNS}. Strong deviations from Curie-Weiss behavior up to $\sim$400$\,$K (inset Fig.~\ref{fig:ACMS}a) \cite{MATSUMOTO,YOSHII07}, imply that these short-range correlations are significant in a wide $T$-range including $T_{CO}$ and may influence the establishing of CO \cite{WEN10}; provided there is a spin charge coupling \cite{XMCD}.\par

Although the AFM/fM metamagnetism presented here may not be resolvable in a majority of LuFe$_{2}$O$_{4}$ samples, the complex phase competition likely has ramifications for all specimens of this material. For example, if disorder, e.g.\ due to oxygen off-stoichiometry, is added to the competing interactions, glassy freezing may be expect to replace long-range spin order at $T_{N}$, as observed in some samples \cite{WANG,PHAN}. Disorder will disrupt most easily the weak inter-bilayer correlations. It is thus natural to expect 3D spin order to be replaced by ``spin-glass-like 2D-ferrimagnetic order'', as reported from early neutron diffraction studies \cite{IIDA}, and for the related YFe$_{2}$O$_{4-x}$ clearly linked to oxygen non-stoichiometry \cite{YFO1}.\par

In summary, we have elucidated the magnetic phase diagram of LuFe$_{2}$O$_{4}$ (Fig.~\ref{fig:Phaselow}) close to $T_{N}$ and determined a ferrimagnetic and an antiferromagnetic spin configuration, which are almost degenerate at $T_{N}$ and $H\!=\!0$. This phase competition is not observed in classical metamagnetic materials, but is remarkably similar to competing CO instabilities at higher $T$. The near-degeneracy arises from geometrical frustration. Together with disorder it can lead for example to glassy freezing instead of long-range order, as observed in many samples, and competing magnetic fluctuations may influence the charge ordering and magnetoelectric coupling.\par

We thank R.P. Hermann, A.B. Harris, J. Voigt, Th. Br\"{u}ckel, and R. Puzniak for useful discussions. Support from the initiative and networking fund of Helmholtz Association by funding the Helmholz University Young Investigator Group ``Complex Ordering Phenomena in Multifunctional Oxides'' is gratefully acknowledged. Work at the SLS was supported by the Swiss National Science Foundation NCCR MaNEP Project. Work at ORNL was supported by the Scientific User Facilities Division, Office of Basic Energy Sciences, US Department of Energy (DoE). MA thanks D. Mandrus, B.C. Sales, W. Tian and R. Jin for their assistance during sample-synthesis, also supported by US-DoE.


\begin{thebibliography}{99}
\bibitem{LOWPHASE}
The transition at $T_{LT}\,\sim$170$\,$K is associated with the development of a strong diffuse magnetic scattering component \cite{WEN10,d0det}, and accompanied by a structural distortion \cite{XU}. The region below $T_{LT}$ will not be discussed here.
\bibitem{AFMPHASE}
Neither an AFM phase nor metamagnetism has been proposed for LuFe$_{2}$O$_{4}$ before, the related YFe$_{2}$O$_{4}$ appears AFM if, and only if, highly stoichiometric \cite{YFO1}.
\end{thebibliography}
\end{document}